# Giant Valley Splitting in Monolayer WS$_2$ by Magnetic Proximity Effect

Tenzin Norden[1, *], Chuan Zhao[1, *], Peiyao Zhang[1], Renat Sabirianov[2, #],

Athos Petrou[1, #], and Hao Zeng[1, #]

[1]Department of Physics, University at Buffalo, the State University of New York, Buffalo, NY 14260, USA;

[2]Department of Physics, University of Nebraska-Omaha, Omaha, NE 68182, USA

*These authors contributed equally to this work

**Abstract**

Lifting the valley degeneracy of monolayer transition metal dichalcogenides (TMD) would allow versatile control of the valley degree of freedom. We report a giant valley exciton splitting of 18 meV/T for monolayer WS$_2$, using the proximity effect from a ferromagnetic EuS substrate, which is enhanced by nearly two orders of magnitude from the 0.2 meV/T obtained by an external magnetic field. More interestingly, a sign reversal of the valley exciton splitting is observed as compared to that of WSe$_2$ on EuS. Using first principles calculations, we investigate the complex behavior of exchange interactions between TMDs and EuS, that is qualitatively different from the Zeeman effect. The sign reversal is attributed to competing ferromagnetic (FM) and antiferromagnetic (AFM) exchange interactions for Eu- and S-terminated EuS surface sites. They act differently on the conduction and valence bands of WS$_2$ compared to WSe$_2$. Tuning the sign and magnitude of the valley exciton splitting offers opportunities for versatile control of valley pseudospin for quantum information processing.

**Introduction**

Proximity effect in condensed matter physics refers to the extension of a particular order parameter of a material into an adjacent material. A well-known example is the superconductivity proximity effect, which occurs when a superconductor is placed in contact with a non-superconducting material [1,2]. Superconductivity emerges in the "normal" region over mesoscopic distances. Another example is the magnetic proximity effect, where a magnetic material induces magnetic moment and magnetic order in a nonmagnetic material adjacent to it [3,4]. As the penetration depth of the magnetic proximity effect is governed by short range exchange interactions, it will be more pronounced in thin film heterostructures, especially in atomically thin 2D heterostructures. Recently, ferromagnetism induced by the interfacial exchange field from a magnetic substrate in graphene and topological insulators has been observed [5-8]. Magnetic proximity effect has also been used to break the time reversal symmetry and lift the valley degeneracy in TMDs [9-11]. Compared to magnetic doping, [12,13] utilizing the magnetic proximity effect allows us to avoid the introduction of defects and reliably separate bulk from surface state effects.

Monolayer TMDs, such as $MoS_2$, $MoSe_2$, $WS_2$ and $WSe_2$ have attracted great interest in recent years because of their unique optical and electronic properties [14-18]. The broken inversion symmetry in monolayer TMDs results in two degenerate yet inequivalent valleys at the corners of the hexagonal Brillouin Zone, labeled as K and K′ valleys. The strong spin-orbit coupling leads to spin splitting of the top valence band and bottom conduction band in the two valleys. The spin orientations are opposite in these valleys, due to the time reversal symmetry [14,15]. This coupling of spin and valley degrees of freedom in monolayer TMDs renders valley dependent optical selection rules, and makes them attractive for spintronics and valleytronics applications [16-19]. For valleytronics applications, creating, switching and detecting valley polarization are required [20,21]. Transient valley polarization has been achieved by optical excitation [16-18] and spin injection [22,23]. Valley pseudospins are useful as qubits since their coherent states can be initialized, controlled, and read out using optical pulses [24]. Lifting the valley degeneracy can lead to robust, non-volatile valley polarization. This has been done by applying an external magnetic field [25-29], as lifting the degeneracy requires breaking the time reversal symmetry. Applying a magnetic field has also been used to rotate the valley exciton pseudospin, a necessity for valley qubits [30]. However, a tiny valley exciton Zeeman splitting of 0.2 meV/T makes valley control difficult at moderate fields. It was found that the valley pseudospin can be rotated by up to 35° at a very large external field of 25 T [30]. Magnetic doping in TMDs to lift the valley degeneracy has also been attempted [12,13,31]. However, valley splitting by doping has yet to be demonstrated. Recently we have shown that, by exploiting the magnetic proximity effect from the ferromagnetic insulator EuS, valley splitting of monolayer $WSe_2$ can be enhanced by more than an order of magnitude [11]. X. Xu et al. also demonstrated valley splitting and polarization in $WSe_2/CrI_3$ [32]. Both of these results show the possibility to control the valley degree of freedom in a feasibly low magnetic field using magnetic proximity effect.

One of the key differences between exchange induced valley splitting by magnetic proximity effect and Zeeman splitting due to an external field is that while Zeeman splitting is nearly identical in value for different TMDs (~ 0.2 meV/T) due to their similar electronic structures and the same orbital moment contributions; the exchange valley splitting by magnetic proximity effect, on the other hand, is governed by the strength of the exchange interactions and is thus tunable depending on the types of TMDs and magnetic substrates, their spatial separation and band alignment. Furthermore, unlike Zeeman splitting which always lowers (raises) the energy of spin up (down) states, exchange interactions can be either positive (FM) or negative (AFM), depending on the nature of the interface and interatomic spacing. A natural question arises is then whether one can exploit this aspect of the exchange interaction to tune both the magnitude and sign of valley splitting of TMDs, which is otherwise unattainable by Zeeman effect. To this end, we chose two types of TMDs, namely $WS_2$ and $WSe_2$ with different band gaps and fabricated their heterostructures with ferromagnetic EuS. $WS_2$ and $WSe_2$ are expected to possess different band alignment with EuS and thus different exchange interactions. We then investigated their valley exciton splitting by magneto-reflectance measurements. We show that in $WS_2$/EuS, the magnetic proximity effect results in a giant valley exciton splitting, up to two orders of magnitude higher than that obtained from an external magnetic field. Moreover, a sign reversal of the splitting is observed by comparing the behavior of $WS_2$/EuS vs $WSe_2$/EuS. Using first principles calculations, we elucidate the critical role of the interface, in particular the effects of the competition between different surface terminations on band alignment and the exchange. The possibility to tune not only the magnitude but also the sign of valley splitting offers opportunities to explore new physics, and provides flexibility in valley control for applications in information processing.

**Results and discussion**

Monolayer $WS_2$ and $WSe_2$ were grown by sulfurization and selenization of $WO_3$ thin layers grown by electron beam deposition, following our previously published procedures [33]. They were then transferred onto magnetic EuS and nonmagnetic Si/SiO$_2$ substrates [34]. The morphologies were studied by optical, scanning electron and atomic force microscopies. Raman and photoluminescence spectroscopies measured at room temperature were used to confirm their monolayer character.

Fig. 1(a) shows the optical microscope image of the as-grown monolayer $WS_2$ on sapphire substrate (the inset is a SEM image). As shown in the image, each single crystal is of regular triangular shape with a size of 10-20 μm. Fig.1(b) and Fig.1(c) are the corresponding optical and SEM images of the films after being transferred onto $SiO_2$ and EuS substrates, respectively. We can see that the transfer process does not lead to change in morphology. In Figure 1(d), (e) and (f), Raman spectra of the WS2 films on sapphire, $SiO_2$ and EuS substrates are shown. For the sample on sapphire substrate, the $E_{2g}$ and $A_{1g}$ phonon features have Raman shifts of 355.3 and 414.7 $cm^{-1}$. The separation between these two peaks is 59.4 $cm^{-1}$, consistent with reported value for monolayer $WS_2$. After being transferred onto $SiO_2$ and EuS substrates, the separation changes to 62.5 and 62.4 $cm^{-1}$, which may be associated with the chemical bonding between the $WS_2$ and $SiO_2$ and EuS substrates. The PL peak energy is at 626nm (1.98 eV) (Fig. 1(g)), which can be attributed to the A excition transition. The PL peak values change slightly to 616.4 (2.01 eV) and 620.9 (2.00 eV) for samples on $SiO_2$ and EuS substrates (Fig. 1(h) and 1(i)). These values are consistent with values reported previously for monolayer $WS_2$.

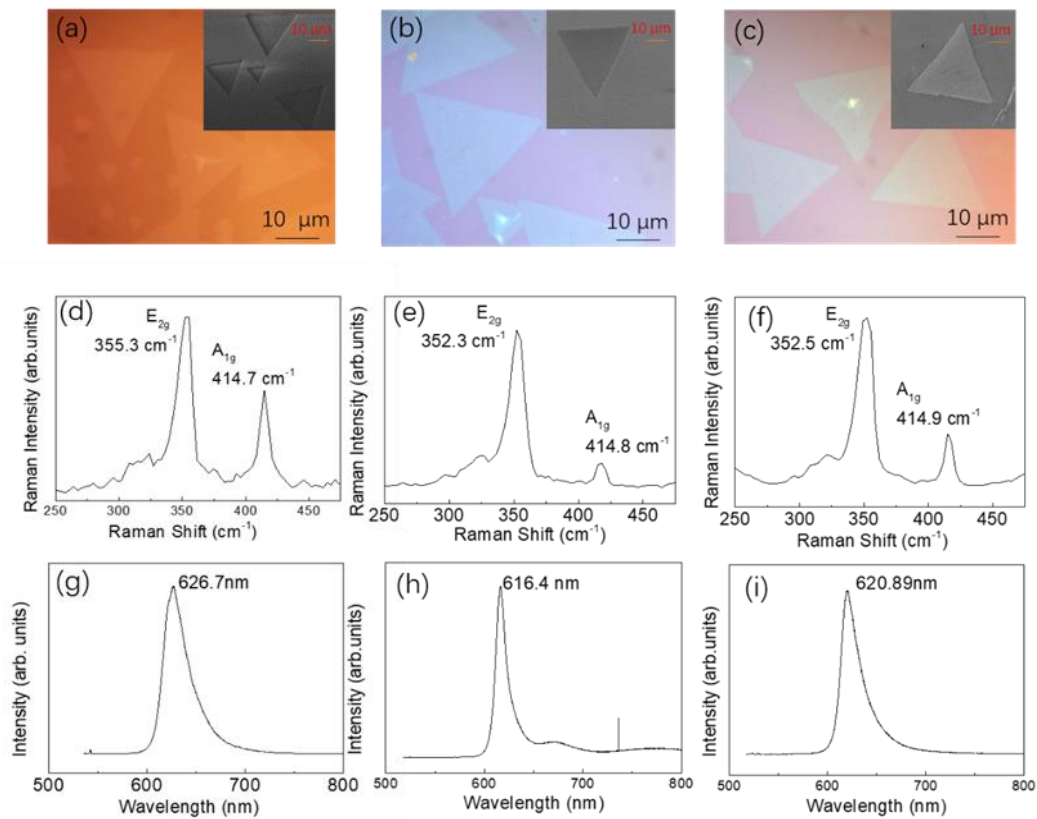

**Figure 1.** Optical microscope images and SEM images (insets) of monolayer $WS_2$ on sapphire substrate (a), $SiO_2$ substrate (b) and EuS substrate (C). Room temperature Raman and PL

spectra of WS$_2$ on sapphire (d and g), SiO$_2$ (e and h) and EuS (f and i).

Temperature dependent magneto-reflectance measurements were performed in the Faraday geometry to determine the valley exciton splitting of TMDs. In this geometry, the optical beam is parallel to the magnetic field and perpendicular to the sample. A linear polarizer and a Babinet-Soleil compensator was used to generate left and right circularly polarized light (see Supporting Information, SI for details).

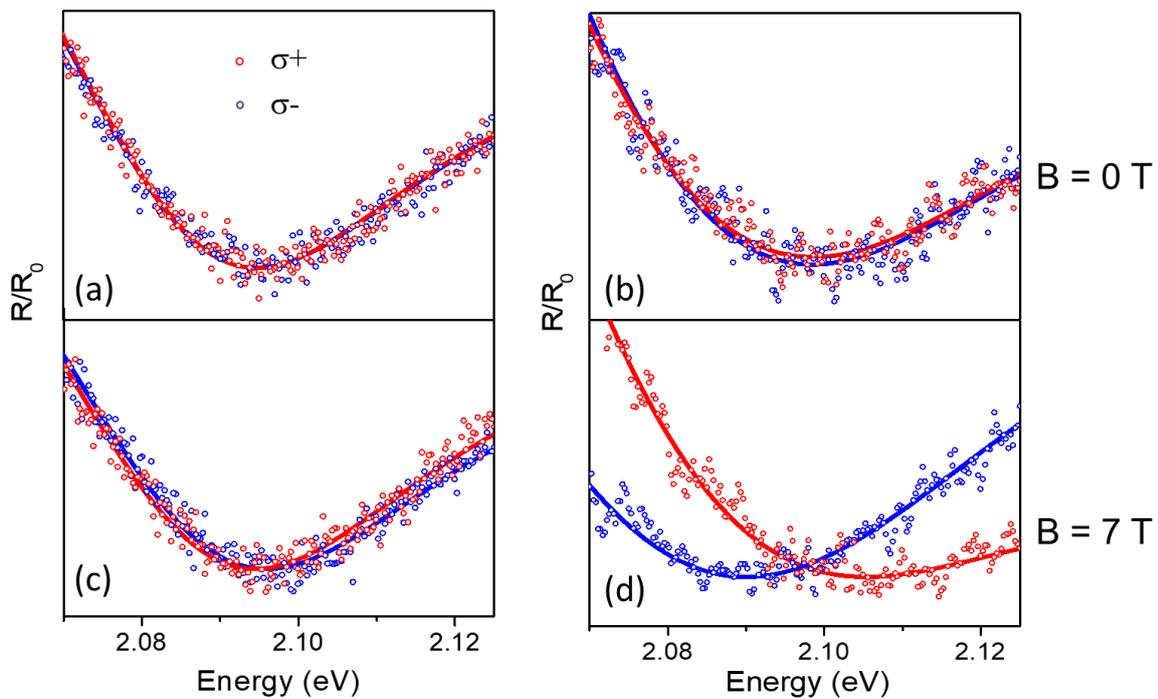

**Figure 2.** Reflectance spectra of monolayer WS$_2$ on Si/SiO$_2$ substrate at B= 0 T (a) and 7 T (c); and on EuS substrate at B= 0 T (b) and 7 T (d).

Figure. 2(a) and 2(c) show the reflectance spectra of monolayer WS$_2$ on Si/SiO$_2$ substrate, measured at 7 K and a magnetic field of 0 and 7 T, respectively. While Fig. 2(b) and 2(d) show the reflectance spectra of WS$_2$ on ferromagnetic EuS substrate measured at identical conditions. The vertical axis labeled as R/R$_0$ represents the ratio between the reflectance of WS$_2$ and either Si/SiO$_2$ or EuS substrate background. A complex (absorptive + dispersive) Fano line shape was used to fit the reflectance spectra (SI) to extract the transition energies (see SI). The local minima were found to be ~ 2.1 eV, corresponding to the A exciton transition energy. In all figures, the left (right) circularly polarized light, denoted as σ+ (σ-), corresponds to the inter-band transition at the K (K′) valley. As can be seen from the top panels in Fig 2(a) and 2(b), at zero field, the σ+ and σ- spectra match well with each other, confirming the degeneracy of the two valleys. When a +7 T magnetic field is applied, however, an energy shift of the exciton transition is observed for WS$_2$ on both Si/SiO$_2$ and EuS substrates (Fig. 2(c) and 2(d)). The σ+ spectrum shifts to lower energy, while the σ- spectrum shifts to higher energy for WS$_2$ on Si/SiO$_2$. While spin, atomic orbital and valley orbital magnetic moments all contribute to the Zeeman shift of the valley states [25-29], it is understood that the valley exciton splitting, which is the difference between the exciton transition energies in the K and K' valleys defined as $\Delta E \equiv E(\sigma-) - E(\sigma+)$, is dominated by the atomic orbital moment of the top valence band at the K and K' valleys, each contributing 2 μ$_B$ and -2 μ$_B$, respectively [14,25,26,28,29]. The Zeeman splitting of the valley exciton is thus $\Delta E = 4\mu_B B$, where B is the applied magnetic field. The valley splitting of WS$_2$ on Si/SiO$_2$ is measured to be 1.5 meV at 7 T, consistent with the prediction and the values reported earlier [28,29]. For WS$_2$ on EuS, however, unexpected behaviors are observed. As can be seen from Fig. 2(d), the separation between the σ+ and σ- spectra is much larger, *i.e.* a much larger valley exciton splitting. Moreover, the sign of valley exciton splitting is reversed, with σ+ spectrum shifting to higher energy while σ- shifting to lower energy. The sign of $\Delta E$ is also opposite to that of WSe$_2$/EuS reported by us previously [11]. This sign reversal suggests interesting physics associated with the magnetic proximity effect that are qualitatively different from the Zeeman effect.

To elucidate the origin of the valley exciton splitting induced by the EuS substrate, we measured and plotted in Fig. 3(a) $\Delta E$ of A exciton transition as a function of the applied magnetic field for monolayer $WS_2$ on EuS and $SiO_2$, resΔΔpectively. We also plot the data of monolayer $WSe_2$ on EuS for comparison There are clear contrasting behaviors in the dependence of $\Delta E$ on B. For $WS_2$ on $SiO_2$, $\Delta E$ increases linearly with increasing magnetic field at a slope of ~ 0.2 meV/T (Fig. 3(a), red dots), consistent with a g-factor of ~ 4. As for $WS_2$ on EuS, $\Delta E$ vs B shows prominent nonlinear behavior, with an initial slope of -18 meV/T at |B| <1 T (for $WSe_2$ on EuS, the slope is +2.5 meV/T). The magnitude of $\Delta E$ for $WS_2$/EuS corresponds to a g-factor of -360, which is nearly two orders of magnitude higher than that on $Si/SiO_2$ substrate and 7 times higher than the value of $WSe_2$/EuS. This giant enhancement in valley exciton splitting clearly originates from the interactions between the monolayer $WS_2$ and the magnetic EuS substrate.

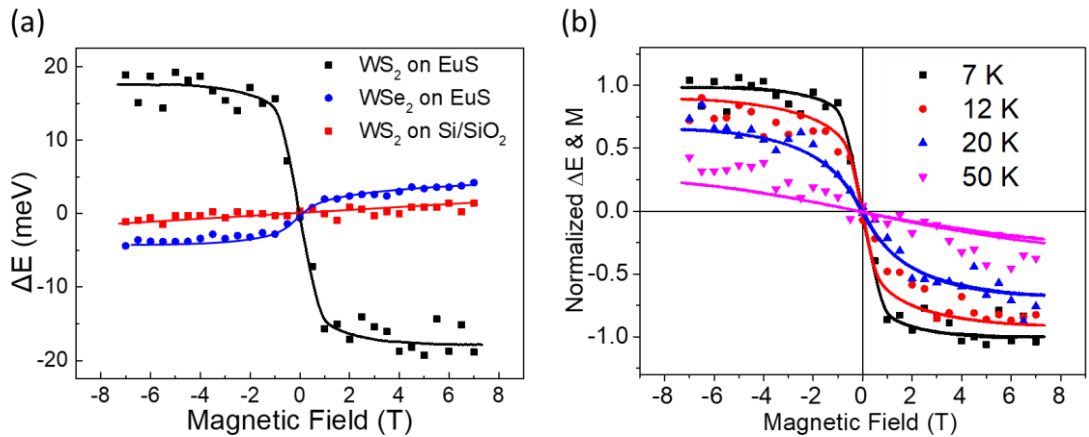

**Figure 3.** (a) Field dependent valley exciton splitting $\Delta E$ of $WS_2$/EuS (black squares), $WS_2$/$SiO_2$ (red squares) and $WSe_2$/EuS (blue circles) measured at 7 K. The lines serve as guide to the eye. (b) Field-dependent valley splitting $\Delta E$ of $WS_2$/EuS and magnetization M of EuS measured in 7K, 12K, 20K and 50K superimposed on each other. Both $\Delta E$ and M are normalized by their saturation values at 7K. Points, normalized $\Delta E$; lines, normalized M.

If the valley splitting originates from the magnetic proximity effect, $\Delta E$ of WS$_2$ as a function of B field should follow the trend of the out-of-plane magnetization of EuS, as $B_{ex} \propto <S_z>$ [11]. To verify this, the field dependent valley exciton splitting of WS$_2$/EuS and the out-of-plane magnetic hysteresis loops of the EuS film were measured at identical temperatures ranging from 7 to 50 K, and plotted in Fig. 3(b). A linear background of 0.2 meV/T is subtracted to reveal the net contribution from MEF, which is denoted as $\Delta E_{ex}$. $\Delta E_{ex}$ and the magnetization M are normalized by their saturated values at 7 K, respectively, and the sign of the magnetization is inverted so that the two sets of data can be conveniently compared. EuS is a soft magnetic material with its easy axis in the plane. As can be seen from Fig. 3(b), the hysteresis loop shows negligible remnant magnetization, and saturates at approximately 2 T at 7 K. With increasing temperature, the saturation magnetization decreases accordingly. As T increases to 50 K, M becomes approximately linear. EuS has a Curie temperature $T_C$ of 16.7 K, and is expected to be paramagnetic at 50 K. However, the EuS film at 50 K shows a magnetic susceptibility larger than that expected for a typical paramagnetic material, suggesting some residual ferromagnetic correlation. As for valley exciton exchange splitting, at 7 K, $\Delta E_{ex}$ increases rapidly with increasing field for |B|<1T; above 1 T, $\Delta E_{ex}$ increases slowly and then tends toward saturation at higher fields. With increasing temperature, the saturated value of $\Delta E_{ex}$ decreases. As T further increases to 50K, $\Delta E_{ex}$ shows approximately linear field dependence with a slope of –0.4 meV/T. As the non-magnetic background has already been subtracted, this value reflects the contribution from MEF of the EuS substrate. The field-dependence of $\Delta E_{ex}$ is observed to be closely mimicking the behavior of magnetization at different measurement temperatures, as can be seen from Fig. 3(b). This unambiguous correlation between valley exciton splitting for WS$_2$ and out-of-plane magnetization of EuS clearly establishes the MEF as the origin of valley splitting. The sign reversal of valley splitting, however, reflects the complicated behavior of exchange interactions, which remains to be understood. Below we show calculation results of electronic structure for WSe$_2$/EuS and WS$_2$/EuS, and extract band edge energies at K and K′ points of the Brillouin Zone using density functional theory, to explain the observed intriguing behavior of valley exciton splitting.

Exchange coupling between the EuS substrate and TMDs occurs predominantly between states of transition metal (Mo, W) and magnetic surface atom (Eu in our case). The exchange interactions are of indirect exchange type mediated by non-magnetic chalcogen elements. The situation is different from the conventional super-exchange picture where exchange is considered between two transition metal cations mediated by a non-magnetic anion. Eu possesses large localized f-state magnetic moment. Its d-orbitals (partially populated with ~ 0.4 e) are relatively weakly polarized (~0.1 μ$_B$) by the core f-electrons and, thus, the exchange is indirect. Nevertheless, the magnitude and sign of exchange interactions depend sensitively on the interface structure, *i.e.* interatomic distances and termination of the surface.

Due to the hexagonal structure of TMDs, it is not possible to construct a perfectly lattice matched interface between TMD and EuS. For example, (100) and (011) surfaces of cubic systems do not match by symmetry, while (111) surface of EuS does not match due to large difference in lattice parameters. Furthermore, (111) surface for EuS and many other cubic systems is polar, and reconstruction inevitably occurs to avoid the polar catastrophe. In most earlier theoretical calculations, a (111) polar surface was assumed [9,10]. This leads to strong electrostatic interaction between TMD and the magnetic substrate, reducing the interlayer distance and promoting the exchange interactions. The calculated valley exciton exchange splitting was often orders of magnitude larger than experimental observations [9,10].

The realistic surface of an EuS thin film may not have a (111) surface terminated simply by either Eu or S because of the polarity issue. Thus, the surface is expected to have about the same number of surface Eu and S sites (which is true for (100) and (110) high symmetry terminations). The TMD deposited on EuS in this case will have nearly the same distance to the surface S and Eu sites. For monolayer WSe$_2$ or WS$_2$, the exchange coupling between W and Eu sites will then be mediated either by single S/Se site in TMD or double chalcogenide "bridge" as shown in Fig. 4. If indirect exchange favors AFM coupling for a single chalcogenide bridge, the interaction for the double chalcogenide bridge would be FM and vice versa. In considered layered systems, exchange interactions should depend strongly on the surface atom arrangements and interlayer distance, as well as surface polarity. W has less than half-filled 4d shell and should have FM indirect exchange and its magnitude is expected to decrease with increasing distance. The average of the exchange effects from Eu- and S-terminated surfaces should give reasonable expectation values of the net substrate effects.

**Structural model**

The structural model constructed for EuS (111) polar surface have relatively low mismatch with monolayers of WS$_2$ and WSe$_2$, which have lattice constants of 3.148 Å and 3.316 Å, respectively. The lattice parameter of EuS is 5.97 Å and it has rock salt structure, making the in-plane lattice constant of (111) surface of EuS ~8.44 Å. Using $\sqrt{3} \times \sqrt{3}$ construction of unit cell we can reduce the lattice constant to ~7.3 Å. Because TMDs are not expected to have large strain in the experimentally deposited systems, we have kept lattice parameter of the supercell equal to equilibrium lattice parameters of WS$_2$ or WSe$_2$. This induces relatively large strain in EuS lattice; however, it is sufficient to explore the effect of the substrate on the electronic states of TMDs. We considered Eu-terminated and S-terminated EuS surfaces that is shown schematically in Figure 4.

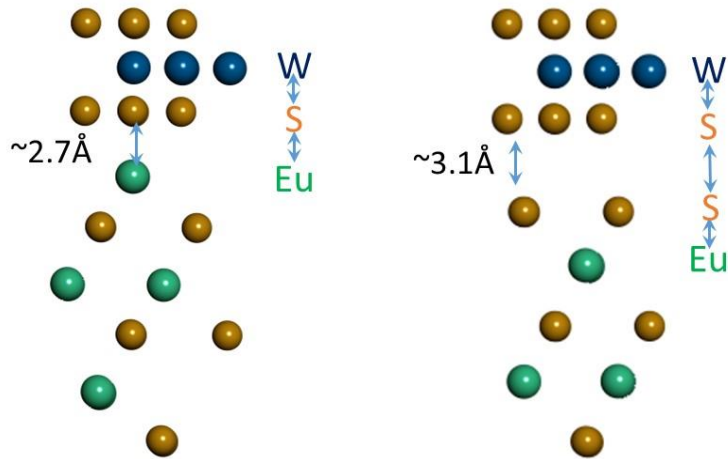

**Figure 4.** Schematic diagram of the structural model of WS$_2$/EuS for Eu-terminated (left) and S-terminated (right) surfaces of EuS.

Figure 4 shows that the exchange field effect between magnetic Eu- and W- states is mediated by the non-magnetic chalcogen element. For WS$_2$, the interaction between Eu and W states is mediated by S for Eu-terminated surface (single S bridge), while Eu-S-S-W (double S-S bridge) is present for the case of S-terminated surface. For WSe$_2$, the interaction between Eu and W states is mediated by Se for Eu-terminated surface (single Se bridge), while Eu-S-Se-W (double S-Se bridge) is present for the case of S-terminated surface.

**Results of DFT calculations**

Our calculation shows that depending on the termination of EuS surface (Eu vs. S), the Eu-W distance varies significantly (Table 1).

**Table 1.** Calculated equilibrium interlayer distances for polar surfaces of WS$_2$ and WSe$_2$ on EuS.

|  | Eu-terminated (Å) | S-terminated (Å) |
|---|---|---|
| WS$_2$ | 2.76 | 3.19 |
| WSe$_2$ | 2.77 | 3.21 |

In our calculations, we kept the distances between TMD and the surface equal for both Eu and S terminations because the realistic surface should have mixture of S and Eu terminations to remove the polarity. In this case the TMD monolayer will be planar and keep nearly the same distance for either termination. This distance was chosen to be the relaxed spacing calculated for Eu-terminated surface (2.76-2.77 Å), due to the stronger electrostatic interactions between the Eu sites and chalcogen elements in TMDs. Thus, the discussion of the differences and similarities of TMD properties deposited on EuS will be the most consistent.

**Eu terminated (111) polar surface** – Overall, $WS_2$ on the polar Eu-terminated EuS surface shows behavior similar to that of $WSe_2$ [11], as seen in band structure plots in Fig. 5(a) and (c) and projected partial density of states (DOS) in Fig. 6(a) and (c). Due to the electrostatic attraction between the surface Eu and S sites, there is a significant redistribution of the charges in $WS_2$ ($WSe_2$) and dipole moment forms at the interface. The Fermi energy falls into the conduction band and is likely due to the polarity of the interface.

The band structure shows considerable shift in conduction and valence bands due to the presence of the EuS substrate. As can be seen from Fig. 6(a) and 6(c), the conduction band edge of $WS_2$ ($WSe_2$) (red) aligns with that of EuS (black) and causes large exchange splitting between the majority and minority states of TMDs. The on-site exchange splitting is about 0.2-0.3 eV, noticeable in the shift of the DOS seen in Fig. 6(a) and 6(c). Note that one should differentiate the edge of the conduction bands from the W-d states responsible for optical transitions. This can be seen clearly from Fig. 5(a) and 5(c). Labeled by the circle in Fig. 5(a) are the conduction band edge states with large exchange splitting between the majority and minority states. They are dominated by interfacial S states that are hybridized with W-states. On the other hand, the edge of the valence states has substantially smaller exchange splitting due to the different character of the bands that is made up mainly of W d-states (Fig. 5(a) and 5(c)), and there are no interface related states at this energy window. As a result, the interfacial states in the conduction band do not contribute to exciton intensity. The distinguishing feature of such a system is the presence of spin polarization in WS$_2$ (WSe$_2$) resulting in finite magnetic moments of about 0.1 $\mu_B$/f.u. This induced moment is significant and should be detectable by X-ray magnetic circular dichroism.

On the other hand, the bottom of the W d-states in the conduction band responsible for the optical transition is located above the conduction band edge. The spin-resolved bands for W d-states are shown in Fig. 5 in red and blue color for majority and minority states, respectively. These states exhibit a spin state "inversion" between K and K′ high symmetry points of the Brillouin zone and are responsible for exciton optical spectra. The on-site exchange splitting of 0.2-0.3 eV is nearly an order of magnitude larger than the exchange shift of W d-states (~0.04 eV). Note that it is the difference in the exchange shift of W d-states between the K and K′ valleys that is responsible for the experimentally observed valley exciton splitting.

**S terminated (111) polar surface –** For the S-terminated (111) polar surface, the Fermi energy falls into the valence band due to the polarity at the interface opposite to that in Eu- terminated surface. (A reconstructed non-polar surface, *i.e.* a surface with near equal number of S and Eu atoms, will have Fermi energy at the center of the bandgap.) There is still a small magnetic moment (~0.01 $\mu_B$/f.u.) observed in the TMD layer. It is worth to note that the band alignment between WS$_2$ and EuS is now different from that between WSe$_2$ and EuS, due to the difference in the bandgap of the two TMD systems. Bandgap of WS$_2$ is about 1.9 eV, while it is only ~1.3 eV for WSe$_2$. Because the bandgap of EuS of 1.6 eV is in between, in the conduction band, the W d-states of WS$_2$ appears to be inside the interface related bands (Fig. 3(b)). This is in clear contrast to Eu-terminated case, where W d- states are above the interface related S states (Fig. 3(a)). On the other hand, the W d-states of WSe$_2$ are located at the bottom of the conduction band and below the interface related states (Fig. 3(d)). As a result of this band alignment, larger exchange field is expected for $WS_2$ than for $WSe_2$ in the conduction band (see Table 3).

The valence band also show opposite shift of W-d states due to the presence of EuS compared to the one in case of Eu-terminated surface, similar to that of conduction band. Clearly, the magnitude of this exchange shift (~ 0.01-0.03 eV) is smaller both for $WS_2$ and $WSe_2$ than Eu-terminated case.

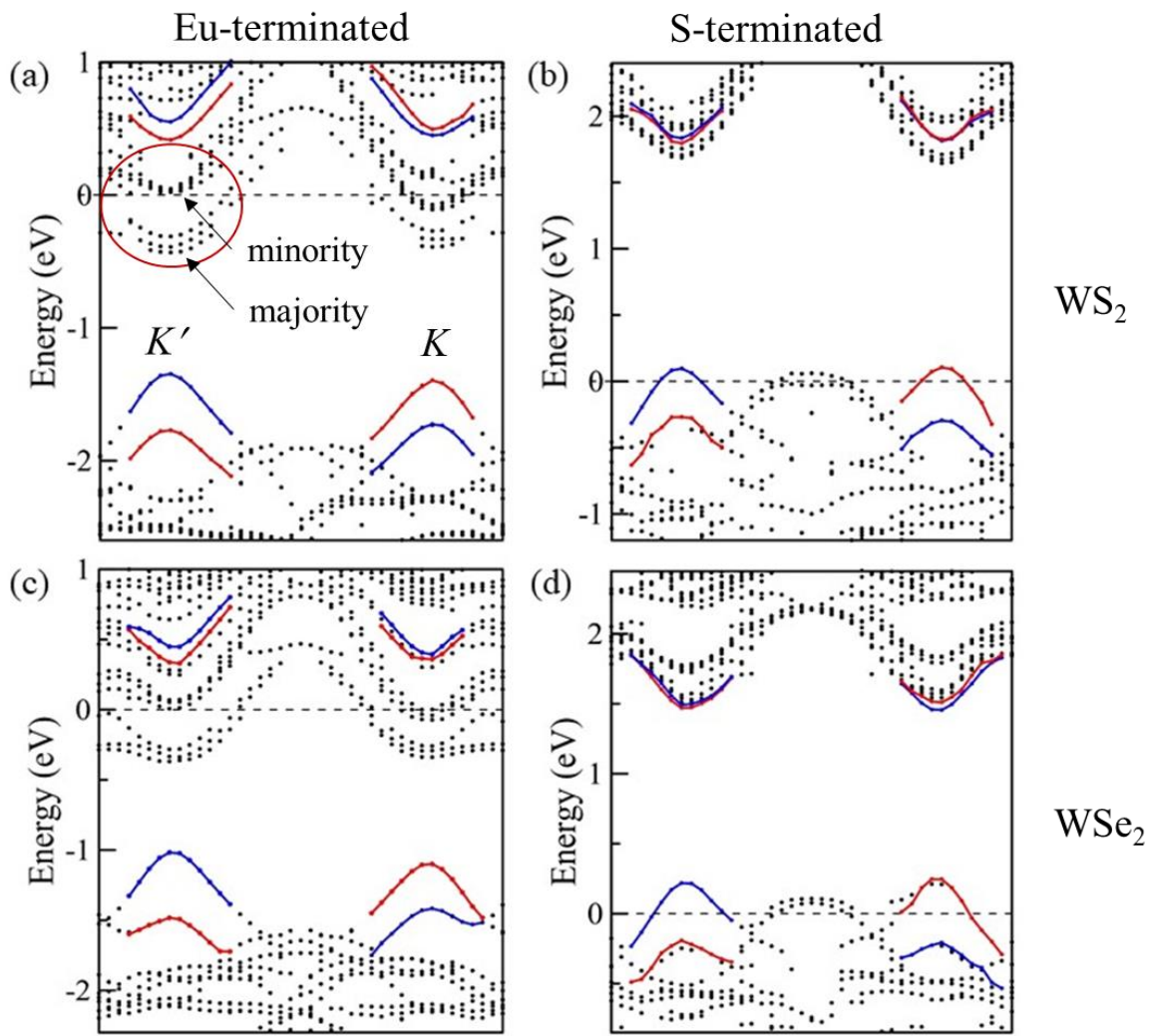

**Figure 5.** Band structure of $WS_2$ (a and b), and $WSe_2$ (c and d). left panel: Eu-terminated EuS, right panel: S-terminated EuS. The W d-states are shown as red (spin-up) and blue (spin-down) bands. Notice that the shifts of the valence and conduction bands in cases of Eu- and S-terminations are in opposite directions due to the opposite polarity.

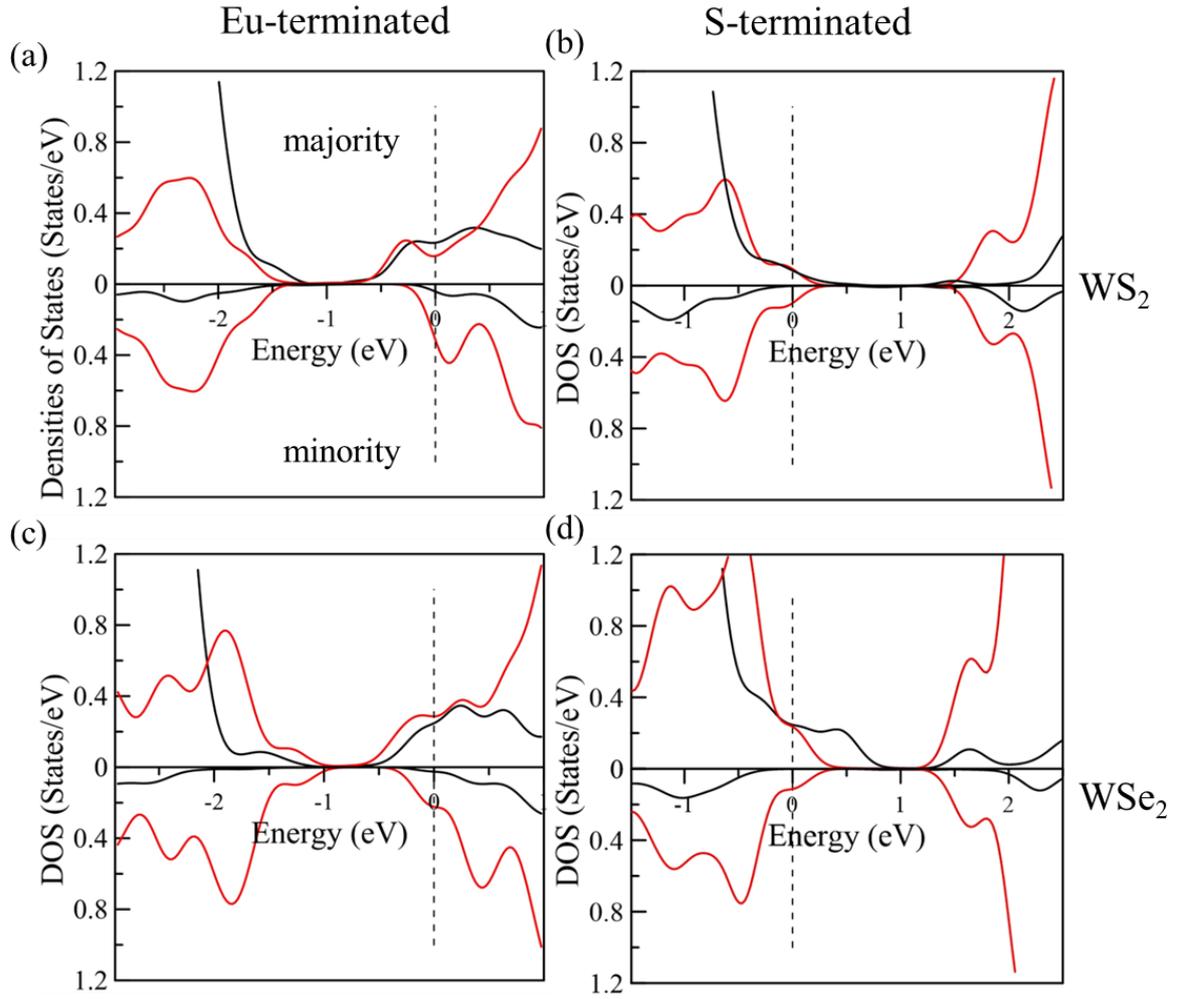

**Figure 6.** Partial Densities of States of surface Eu (black) and W (red) of $WS_2/EuS$ and $WSe_2/EuS$ for polar surface of EuS: (a) $WS_2/EuS$ terminated by Eu atoms; (b) $WS_2/EuS$ terminated by S atoms; (c) $WSe_2/EuS$ terminated by Eu atoms; and (d) $WSe_2/EuS$ terminated by S atoms. (Upper half of the plots -majority, lower half of the plots -minority). Fermi energy is placed at zero and marked by dashed vertical line.

**W d-states energies at K and K′ valleys and valley A exciton splitting**

In the following discussions, we focus on excitonic transitions (A exciton) and valley exciton splitting. Unless otherwise specified, the conduction band and valence band refer to those of W d-states, and the band gap refers to the optical band gap. The energies of the lowest conduction and highest valence bands of TMDs are extracted and displayed in Table 2, to obtain the valley A exciton transition energies assuming vertical excitation at K and K′ points.

**Table 2.** K and K′ valley energies for the top valence and bottom conduction W d-states bands for each spin.

|  | WS$_2$ | | | | WSe$_2$ | | | |
|---|---|---|---|---|---|---|---|---|
|  | Eu-terminated | | S-terminated | | Eu-terminated | | S-terminated | |
|  | K | K′ | K | K′ | K | K′ | K | K′ |
| $E_v(\downarrow)$ (eV) | 0.3859 | 0.7651 | 1.9609 | 2.3535 | 0.2325 | 0.6328 | 1.6153 | 2.0657 |
| $E_v(\uparrow)$ (eV) | 0.7156 | 0.3414 | 2.3620 | 1.9884 | 0.5508 | 0.1669 | 2.0927 | 1.6552 |
| $E_c(\downarrow)$ (eV) | 2.5597 | 2.6619 | 4.0715 | 4.0453 | 2.0618 | 2.0998 | 3.3043 | 3.3412 |
| $E_c(\uparrow)$ (eV) | 2.6065 | 2.5271 | 4.0815 | 4.0920 | 2.0098 | 1.9885 | 3.3584 | 3.3180 |
| $E_g$ (eV) | 1.8909 | 1.8968 | 1.7195 | 1.6918 | 1.4590 | 1.4670 | 1.2657 | 1.2755 |
| $\Delta E_{ex}$ (meV) | 5.9 | | -27.7 | | 8.0 | | 9.8 | |
| $\Delta E_{avg}$ (meV) | -11 | | | | +8.9 | | | |

The exciton energies of TMDs depend on the relative shifts of valence and conduction bands due to the exchange interactions between the TMDs and the magnetic substrate, as shown in Table 2. The valley A exciton splitting, defined as $\Delta E_{ex} \equiv E(K') - E(K)$, for WS$_2$ vs WSe$_2$ on the polar surfaces are quite different. For WS$_2$, it is relatively small and positive for Eu-terminated surface (+5.9 meV), while it is large and negative for S-terminated surface (-27.7 meV). The reconstructed non-polar EuS surface is expected to have near equal ratio of S and Eu sites. Thus, averaging effects from Eu and S terminated surfaces should give reasonable expectation values of the net valley exciton splitting. Accordingly, the calculated valley A exciton splitting for WS$_2$ on EuS is about -11 meV, which is not far from the experimental value of -18 meV. For WSe$_2$ on EuS, on the other hand, the splitting is of similar value and positive for both Eu- and S- terminated surfaces. As a result, the averaged valley exciton splitting is about +9 meV. This is close to the value reported earlier by us considering the Eu-terminated polar surface alone [11]. Thus, by considering the realistic EuS surface and competing exchange effects from the Eu- and S- terminated sites, our theoretical results are consistent with the experimentally observed sign reversal in valley exciton splitting in $WS_2$ compared to $WSe_2$.

**Fitting DFT results to the Model Hamiltonian**

To fully understand the sign reversal in valley exciton splitting, we fitted the DFT results using minimum band model to extract the band gap, spin orbit coupling and exchange parameters. As discussed in our previous studies, the exchange field produces opposite band edge shifts at K and K′ valleys due to spin contributions, as the spin characters of the bands are opposite in different valleys [11]. However, if exchange field would be the same for conduction and valence bands, then neither the spin (because interband optical excitation occurs between states of the same spin) nor the valley orbital moment would contribute to the exciton shift in a magnetic field due to reorientation of substrate magnetization. Thus, atomic orbital moment is expected to contribute to the valley exciton splitting in a manner similar to that from an external field [26-29]. Nevertheless, the separation of the contributions of exchange effects from the spin and orbital moments is not straightforward. Therefore, we will discuss the combined effect as obtained in DFT+U calculations.

The Hamiltonian of TMDs, which has orbital parts, describing two band *k·p* gapped Dirac states with addition of spin-orbit coupling and exchange interactions are:

$$H_{orb} = \hbar v_f(\tau k_x \sigma_x + k_y \sigma_y) + \frac{E_g}{2}\sigma_z \qquad \text{Eq.1}$$

$$H_{SO} = \tau s_z[\lambda_C \sigma_+ + \lambda_V \sigma_-] \qquad \text{Eq.2}$$

$$H_{exch} = -s_z \mu_B (B_z^C \sigma_+ + B_z^V \sigma_-) \qquad \text{Eq.3}$$

Where $v_f$ is the Fermi velocity of the Dirac electrons, $E_g$ is the staggered potential (gap), $\sigma_i$ are the peudospin Pauli matrices operating on the sublattice A and B, $\sigma_0 = \begin{pmatrix} 1 & 0 \\ 0 & 1 \end{pmatrix}$, $\sigma_x = \begin{pmatrix} 0 & 1 \\ 1 & 0 \end{pmatrix}$, $\sigma_y = \begin{pmatrix} 0 & -i \\ i & 0 \end{pmatrix}$, $\sigma_z = \begin{pmatrix} 1 & 0 \\ 0 & -1 \end{pmatrix}$, $\sigma_\pm = \frac{1}{2}(\sigma_0 \pm \sigma_z)$, and $k_x$ and $k_y$ are the Cartesian components of the electron wave vector measured from $K$ ($K'$); parameter $\tau = 1$ (-1) for $K$ ($K'$) valleys. $\lambda_C$ and $\lambda_V$ are intrinsic spin-orbit parameters for the conduction and valence bands, respectively, and $s_z$ is the Pauli spin matrix in the z direction. We have to introduce exchange fields $B_z^C$ and $B_z^V$ separately for the conduction and valence bands, to fit the obtained DFT eigenvalues because the DFT calculations reflect the combined effects and it is not straightforward to separate various contributions into a simple model.

**Table 3.** Parameters of the minimal band model calculated from the DFT band structure results.

| | Δ (eV) | $\lambda_c$ (eV) | $\lambda_v$ (eV) | $B_z^C$ (T) | $B_z^V$ (T) | $\Delta E_{ex} = 2\mu_B(B_z^C - B_z^V)$ (eV) |
|---|---|---|---|---|---|---|
| WS$_2$, Eu-termination | 2.0368 | 0.0908 | 0.3768 | 478 | 428 | +0.0058 |
| WS$_2$, S-termination | 1.9063 | 0.0184 | 0.3831 | -311 | -74 | -0.0276 |
| WSe$_2$, Eu-termination | 1.6442 | 0.0297 | 0.3921 | 777 | 708 | +0.008 |
| WSe$_2$, S-termination | 1.4732 | 0.0387 | 0.4440 | -149 | -233 | +0.0098 |

The fitted band gaps, the spin-orbit parameters and the exchange parameters are shown in Table 3. Depending on terminations, the exchange interaction between the EuS substrate and W is mediated by either a single (in the case of Eu-termination) or double (in the case of S-termination) chalcogen sites. As can be seen from Table 3, the effective exchange fields for Eu-terminated surface sites are larger and positive (FM). The large on-site exchange splitting of Eu d-states (*i.e.* strongly spin polarized) translates to the opposite shift in the majority and minority W d-states with overall large FM exchange (expected for less than half-filled d-shell elements). The magnitude of the effective exchange fields is very large, reaching $B_z^C = 777$ T in the conduction band of WSe$_2$ and resulting in a giant valley splitting of $\Delta E_c$ = 90 meV in the conduction band of WSe$_2$ for Eu-terminated surface. These values are 478 T and $\Delta E_c$ = 55 meV in the conduction band of WS$_2$ for Eu-terminated surface. The effective exchange fields for S-terminated EuS surface sites are smaller and negative (AFM), being -149 T and -311 T in the conduction bands of WSe$_2$ and WS$_2$, respectively. The opposite signs of exchange fields for different terminations result in competing interactions for a realistic reconstructed EuS surface with equiatomic ratio. The calculated exchange fields for the valence bands $B_z^V$ are also displayed in Table 3, showing similar trend to those of $B_z^C$.

One should clearly distinguish between the valley splitting of the conduction and valence bands, and the valley exciton splitting. The valley splitting of the conduction and valence bands is the energy difference of those bands between K and K′ valleys, which depends on the effective exchange fields $B_z^C$ and $B_z^C$. For both WS$_2$ and WSe$_2$, the indirect exchange interactions are larger and FM for Eu-termination, while smaller and AFM for S-termination. As a result, the net effective exchange between TMDs and EuS is expected to be FM for a realistic EuS substrate with surface reconstruction. In the case of the A exciton, the effective exchange field always lowers (raises) the energy of spin up (down) bands in the K (K′) valley (Fig. 7), and the valley splitting of the conduction and valence bands do not change sign for different TMD materials. The valley exciton splitting, on the other hand, is defined as the difference between the exciton transition energies of K and K′ valleys. Its sign is dependent on the relative shift of the conduction and valence bands in the two valleys. For WS$_2$, the effective exchange field for the conduction band $B_z^C$ is smaller than that for the valence band $B_z^V$, i.e. the exchange splitting in the conduction band $\Delta E_c$ is smaller than $\Delta E_v$ in the valence band. The band gap in the K valley $E_{g,K}$ thus increases while $E_{g,K'}$ decreases in the K′ valley, resulting in a negative valley exciton splitting $\Delta E_{ex}$ (see Fig. 7(a)). The situation for WSe$_2$ (Fig. 7(b)) is just the opposite, leading to a positive $\Delta E_{ex}$.

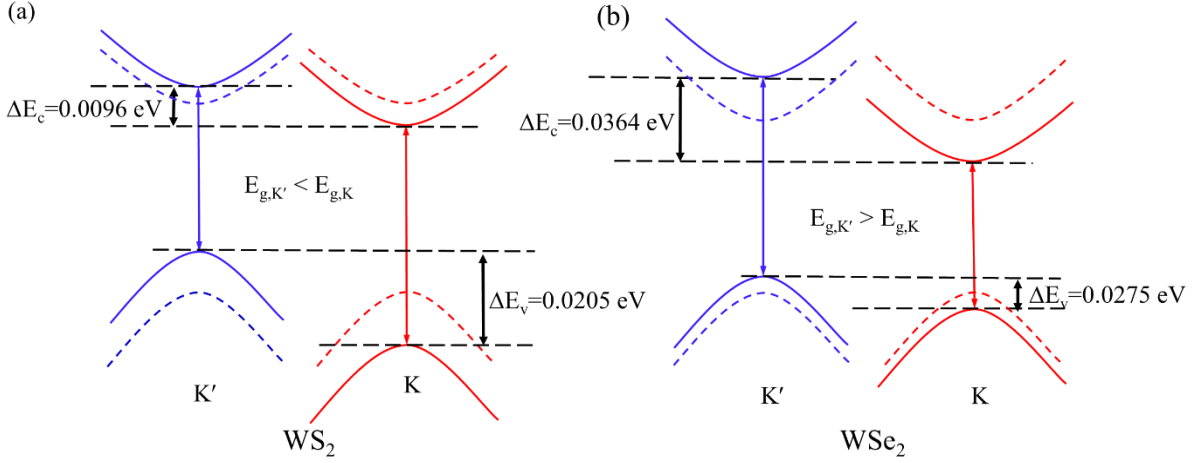

**Figure 7.** A schematic of the splitting of the conduction and valence bands responsible for A exciton transition at K and K′ valleys for WS$_2$ (a) and WSe$_2$ (b). Spin up (down) bands are shown in red (blue). Dashed curves are for degenerate bands without a magnetic substrate; solid curves represent bands of TMDs with EuS substrate. In WS$_2$, $\Delta E_c < \Delta E_v$, leading to a smaller $E_g$ in K′ valley than that in K valley and a negative valley exciton splitting $\Delta E_{ex}$; in WSe$_2$, $\Delta E_c > \Delta E_v$, and $E_g$ in K′ valley is larger than that in K valley, hence a positive $\Delta E_{ex}$.

Experimentally, magneto-reflectance were used to probe the valley exciton splitting. However, it cannot yield information on exchange splitting for the conduction and valence bands separately and thus the type of effective exchange interaction. Additional measurements by, for example, angle-resolved photoemission spectroscopy could give information about the occupied states, while inverse photoemission spectroscopy may shed light on the conduction states. Both experiments can be technologically relevant: while valley exciton splitting allows optical access of valley degree of freedom by light helicity and frequency, valley splitting of the conduction and valence bands allows electric gating to realize valley polarization of either electrons or holes.

From the above discussion, it can be seen that there are two essential factors in controlling the magnitude and sign of the valley exciton splitting in TMDs by magnetic proximity effect: the surface termination of EuS and the band alignment between TMDs and EuS. The indirect exchange interactions are FM for the one mediated by a single chalcogen bridge with Eu-termination, and AFM for the one mediated by a double chalcogen bridge with S-termination. This leads to competition between the two types of exchange interactions for a realistic EuS surface. Moreover, the band alignment determines the magnitude of the exchange splitting of the conduction and valence bands in TMDs. Together, they can determine the sign of valley exciton splitting depending on whether the band gap is widened or narrowed in K and K′ valleys. Such understanding provides important guidance for designing the TMD/Magnetic heterostructures with tunable valley splitting. For example, surface termination can be varied by controlling the growth orientation of single crystalline EuS using molecular epitaxy; band alignment can be tuned by different combinations of magnetic materials and TMDs; and alloying of different TMDs such as $W(S_xSe_{1-x})_2$ can lead to continuous tuning of the magnitude and sign of valley exciton splitting [35]. Giant enhancement in valley exciton splitting, together with tunable magnitude and sign, would allow versatile control of valley pseudospin, enabling its arbitrary rotation and switching, and possibly entangled pseudospins with controllable phases. Most recently, several groups reported discovery of intrinsic 2D ferromagnetic materials [36-38]. Combining these 2D magnets with 2D TMDs may offer a practical approach for emerging valleytronics applications.

**Conclusion**

Magnetic proximity effect from a ferromagnetic EuS substrate results in a giant valley exciton splitting in monolayer $WS_2$. A sign reversal from positive for $WSe_2$ to negative for $WS_2$ is also observed. This is attributed to the competing exchange interactions for Eu- and S- termination for a realistic EuS substrate with surface reconstruction, together with different band alignments between TMDs and EuS. The ability to tune the magnitude and sign of valley splitting allows convenient control of valley pseudospin for quantum information processing.

## Methods

### Sample preparation

Monolayer TMDs, including $WSe_2$ and $WS_2$, were prepared by selenization or sulfurization of electron-beam evaporated ultrathin transition-metal-oxide films on sapphire substrates. Details are in supporting information.

### Film transfer

The as-grown monolayer TMD films were transferred onto $Si/SiO_2$ and EuS substrates by modified published procedures [34]. Briefly, monolayer $WSe_2$ or $WS_2$ on a sapphire substrate was spin-coated by polystyrene (PS). After 5 min baking at 120°C, a water droplet was placed on the PS surface. The sample's edge was then poked by tweezers, and the water penetrated between the film and the substrate. After several minutes, the film completely separated from the substrate and floated on the water surface. The film was then transferred onto a $Si/SiO_2$ or EuS substrate, followed by 5 min baking at 80 °C. The baking of PS at sufficiently high temperatures will eliminate wrinkles in the TMDs. The PS was removed by immersing the sample in toluene for 15 min. After repeated cleaning, the sample was then annealed in an ultrahigh vacuum chamber at 350 °C for 180 min to remove any PS residual and improve the interface quality.

### Optical measurements

For magneto-reflectance measurements, the samples were placed on the cold finger of a continuous-flow optical cryostat operated in the 5–300 K temperature range. The cryostat was mounted on a three-axis translator with a spatial resolution of 10 μm in each direction. The *x*- and *y*-translation stages allow us to access a single TMD crystal. The cryostat tail was positioned inside the room temperature bore of a 7 T superconducting magnet. A collimated white-light beam was used for the reflectivity work. The incident light was focused on the sample using a microscope objective with a working distance of 34 mm. The incident beam was polarized either as σ+ or σ− using a Babinet–Soleil compensator. The objective collected the reflected beam from the sample in the Faraday geometry and the light was focused onto the entrance slit of a single monochromator that uses a cooled charge-coupled device detector array.

**Magnetization measurements**

The field-dependent magnetization of EuS at different temperatures was measured by the VSM option of a quantum design physical property measurement system. The magnetic field was applied in the direction perpendicular to the film plane, and thus only the out-of-plane component of the magnetic moment was measured.

**Computational details**

Density-functional based calculations are performed by using projector augmented wave (PAW) method (as implemented in VASP) within GGA-PBE approximations. van der Waals interaction is taken into consideration using DFT-D3 method. Electron wave function cut off energy is 400 eV. 11×11×1 Γ-centred Monkhorst-Pack grids was used for Brillouin-zone integration. A vacuum of about 15Å separates periodically repeated slabs. Structural relaxation is carried out using the conjugate-gradient algorithm until the Hellmann-Feynman force on each atom is less than 0.01 eV/°A, respectively. We used Hubbard U on f-states of Eu $U = 7.5$ eV, $J = 0.6$ eV, as well as $U = -4$ eV on W d-states to align the band centers of TMS and substrate.

**Acknowledgements**

Financial support by the US National Science Foundation DMR 1229208, DMR-1104994 and CBET-1510121is gratefully acknowledged.

**Author information**

**Affiliations**

Tenzin Norden, Chuan Zhao, Peiyao Zhang, Athos Petrou1, and Hao Zeng

Department of Physics, University at Buffalo, the State University of New York, Buffalo, NY 14260, USA

Renat Sabirianov

Department of Physics, University of Nebraska-Omaha, Omaha, NE 68182, USA

**Contributions**

H.Z. and A.P. conceived, designed and guided the experiments. T.N., C.Z., and P.Z. performed the experiments and data analysis. R.S. performed the first-principle calculations. H.Z. and R.S. wrote the manuscript. All the authors commented on the manuscript.

**Competing interests**

The authors declare no competing interests.

**Corresponding authors**

Correspondence to R. Sabirianov, rsabirianov@mail.unomaha.edu, or A. Petrou, petrou@buffalo.edu, or H. Zeng, haozeng@buffalo.edu

**Supporting Information**

**Giant Valley Splitting in Monolayer WS$_2$ by Magnetic Proximity Effect**

Tenzin Norden[1], Chuan Zhao[1], Peiyao Zhang, Renat Sabirianov[2],

Athos Petrou[1], and Hao Zeng[1]

[1]Department of Physics, University at Buffalo, the State University of New York, Buffalo, NY 14260, USA;

[2]Department of Physics, University of Nebraska-Omaha, Omaha, NE 68182, USA

**Magneto-reflectance measurements**

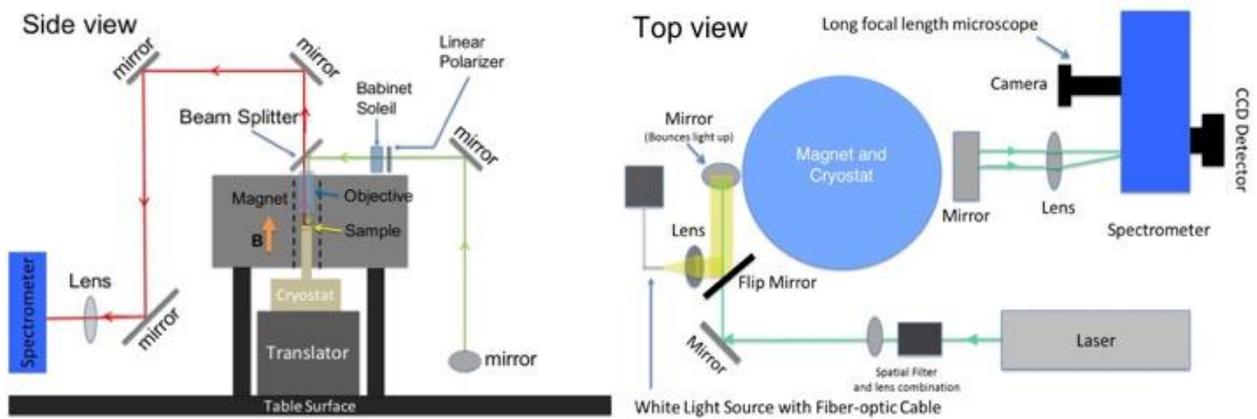

A schematic of the setup for magneto-reflectance measurements. Positive magnetic field is defined as the upward direction. The incident light is right and left circularly polarized by a combination of a babinet soleil and a linear polarizer. The reflected light from the sample is collected using a microscope objective and the light was then focused onto the entrance slit of a single monochromator that uses a cooled charge-coupled device detector array.

**Magneto-reflectance curve fitting**

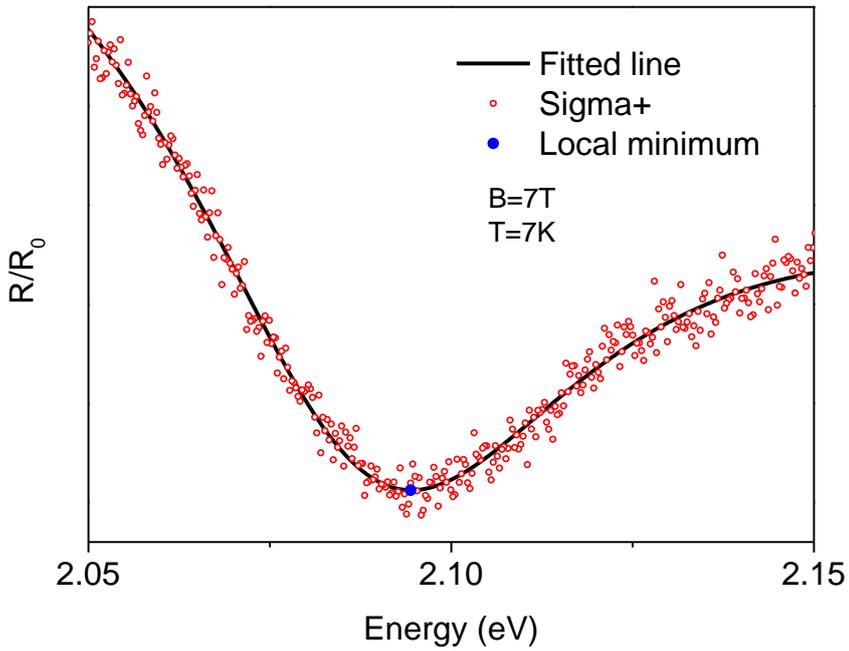

For reflectance signal, it is conventional to use absorptive and dispersive line shape in addition to a linear function to fit the curve and use the local minimum as the **peak position**. As shown in figure above, the blue dot is the fitted peak position.

The fitting equation is:

$$f(x) = A \frac{(q\frac{m}{2} + x - \mu)^2}{(\frac{m}{2})^2 + (x - \mu)^2} + kx + b,$$

here $A$ is the amplitude, $q$ is the Fano parameter which represents the ratio of resonant scattering to the background scattering, $m$ is the width of the line shape, $k$ is the linear slope and $b$ is the intercept.